\documentclass [12pt]{article}
\usepackage {amssymb}
\usepackage {amsmath}
\usepackage[dvips]{epsfig}

\begin{document}

\begin{center}
\Large\textbf{Analytical solution of diffusion equation for point
defects}
\\[2ex]
\normalsize
\end{center}

\begin{center}
\textbf{O. I. Velichko}
\end{center}

\begin{center}
\bigskip

% \textit{Belarusian State University of Informatics and
% Radioelectronics, 6, P.~Brovki Str., Minsk, 220013 Belarus}

{\it E-mail address (Oleg Velichko):} velichkomail@gmail.com

\end{center}

\textit{Abstract.} The analytical solution of the equation
describing diffusion of intrinsic point defects has been obtained
for a one-dimensional finite-length domain. This solution is
intended for investigating and modeling the changes in defect
distributions during fabrication of semiconductor devices with
layer-type structures. With this purpose, the Robin-type boundary
conditions were imposed on both edges of the domain. Using the
solution obtained, the calculations of distributions of point
defects for different boundary conditions and different defect
migration lengths have been carried out. For the case of
generation of nonequilibrium point defects due to implantation of
hydrogen ions, the influence of the surface on the concentration
and spatial distribution of nonequilibrium point defects was
investigated depending upon the implantation energy.

\bigskip

% {\it PACS:} 61.72.Cc, 61.72.Tt, 66.30.Dn, 07.05.Tp

{\it Keywords:} silicon; implantation; annealing; diffusion;
modeling

\section{Introduction}

The electrophysical parameters of silicon integrated microcircuits
and other semiconductor devices are determined by the state of a
defect-impurity system of doped regions. Now, submicron regions of
semiconductor devices are formed by means of ions implantation
with the subsequent low-budget thermal annealing. During
annealing, the main fraction of the nonequilibrium defects
generated by ion implantation is eliminated. Because hydrogen
atoms readily passivate dangling bonds, introduction of hydrogen
into silicon substrates can be used for further improvements in
the device performance due to decreasing the imperfections of the
crystalline lattice and eliminating undesirable electronic states
from the band gap \cite{Pankove-91}. Introduction of hydrogen can
be carried out by means of silicon treatment in a hydrogen
containing-plasma \cite{Pankove-91,Zhang-02} or due to
implantation of hydrogen ions. In both cases introduction of
hydrogen ions is accompanied by generation of additional defects
in the near surface region. Due to the smallness of hydrogen
atoms, it is found that single point defects, namely, vacancies
and self-interstitials, will be generated in undoped silicon. On
the other hand, after implantation of a high fluence of hydrogen
ions, due to diffusion and quasichemical reactions of generated
point defects among themselves, with hydrogen atoms and other
imperfections of crystalline lattice, the thin heavily damaged
layer, i. e., a quite deep weakened zone, can be formed in the
bulk of a semiconductor. As a result, the active layer with
SiO$_{2}$ isolation can be separated from the rest of the bulk
substrate due to splitting which takes place inside the weakened
zone. In such a way different structures called
silicon-on-insulator (SOI) are formed \cite{Chao-05} which have a
number of advantages in comparison with the electric isolation
fabricated by the traditional technology.

It is worth noting that nonequilibrium point defects can be mobile
even at room temperature. Indeed, according to the temperature
dependence obtained in the paper \cite{Panteleev-76}, the
diffusivity of silicon self-interstitials atoms $d_{i}^{I}$ = 1.06
$\mu$m$^{{\rm 2}}$/s for a temperature of 300 K. On the other
hand, it follows from the data of \cite{Hallen-98} that this
diffusivity is equal to 3.2$\times $10$^{{\rm 4}} \mu $m$^{{\rm
2}}$/s. The characteristic diffusion length of silicon
self-interstitials $L^{I} = \sqrt {d_{i}^{I} t}$ obtained for
these values of diffusivity varies from 3.26 $\mu $m to 566 $\mu
$m for the time duration $t$ = 10 s. It means that even at room
temperature silicon self-interstitials diffuse easily far away
from the boundaries of active regions. Thus, distributions of
nonequilibrium point defects in fabricated semiconductor devices
are determined not only by their generation in the local domains,
but also by defect redistribution due to diffusion.

For calculation of distributions of point defects in the paper by
Minear et al. \cite{Minear-72} the analytical solution of the
equation

\begin{equation} \label{Minear}
d_{i} \,{\frac{{d^{\, 2}\,C^{D}}}{{d\,x^{2}}}} -
{\frac{{C^{D}}}{{\tau _{i}} }} + G^{R}(x) = 0  \qquad
\end{equation}

\noindent describing diffusion of point defects was obtained on
the semiinfinite interval [0,+$\infty $]. The case of the constant
coefficients $d_{i} $ and $\tau _{i} $ was considered. Here $C^{D}
= C^{D}\left( {x} \right)$ is the concentration of point defects;
$d_{i}$ and $\tau_{i}$ are the diffusivity and the average
lifetime of point defects in an intrinsic semiconductor,
respectively.

It was supposed in \cite{Minear-72} that nonequilibrium point
defects were continuously generated during ion implantation of
impurity atoms and diffused to the surface and into the bulk of a
semiconductor. The surface was considered to be a perfect sink for
point defects. The concentration of nonequilibrium defects was
also set equal to zero at infinity. It was supposed that the
generation of nonequilibrium point defects is determined by two
factors, namely, generation due to the primary Rutherford
scattering and secondary cascades and generation by hard-sphere
interaction at or near the end of ion's track. Then, the total
generation rate of point defects in the volume unit can be
approximated by an expression with two  summands:

\begin{equation} \label{Minear-Gen}
 G^{R}(x) = G_{m}^{Ruth}\, \mathrm{erfc} \left( { - {\frac{{x - R_{p}}
}{{\Delta R_{p} }}}} \right) + G_{m}^{R} \exp {\left[ { -
{\frac{{\left( {x - R_{p}} \right)^{\,2}}}{{2\Delta R_{p} ^{\,
2}}}}} \right]} \, {\rm ,}
\end{equation}

\noindent where $G_{m}^{Ruth}$ and $G_{m}^{R}$ are the maximal
values of point defect generation rates; $R_{p}$ and $\Delta
R_{p}$ are the average projective range of implanted ions and
straggling of the projective range, respectively.

A similar solution was obtained in \cite{Ryssel-86} for the Robin
boundary condition on the surface of a semiconductor:

\begin{equation} \label{Ryssel}
- {\left. {d_{i} \,{\frac{{d\,C^{D}}}{{d\,x}}}} \right|_{x = 0}}
 + {\rm v}^{S}C^{D}(0) = 0
 \ {\rm ,}
\end{equation}

\noindent where ${\rm v}^{S}$ is the parameter describing the
velocity of point defect trapping on the surface of a
semiconductor. Only the second term in the right-hand side of
expression (\ref{Minear-Gen}) is used for the generation rate of
nonequilibrium defects.

At present, in the modern silicon technology, different layered
structures such as Si$_{{\rm 1}{\rm -} {\rm x}}$Ge$_{{\rm x}}$/Si
\cite{Portavoce-04,Leitz-06,Bhandari-08} and silicon-on-insulator
\cite{Chao-05} are widely used. Therefore, it is reasonable to
obtain an analytical solution of the equation for diffusion of
intrinsic point defects in a finite-length domain $[0,x_{B}]$. The
solution obtained can be helpful for studying the form of point
defect distributions under characteristic conditions used in
processing semiconductor substrates and for verification of
numerical solutions. This solution can be also applied for
modeling a number of the processes of diffusion of vacancies and
silicon self-interstitial because the parameters describing the
transport processes of point defects in silicon and known from the
literature differ by many orders of magnitude \cite{Pichler-04}.

\section{The boundary value-problem for defect diffusion}

The diffusion equations for vacancies and silicon
self-interstitials that take into account different charge states
of intrinsic point defects and drift of the charged species in the
built-in electric field were obtained in
\cite{Velichko-84,Velichko-88}. These equations have the following
form:

\textbf{1) equation of vacancy diffusion}

\begin{equation} \label{Vacancy}
\begin{array}{l}
 \nabla {\left[ {d^{V}\left( {\chi}  \right)\,\nabla \tilde {C}^{V\times} }
\right]} - \nabla {\left[ {\left( {\omega ^{\chi}  - 1}
\right){\displaystyle \frac{{\partial \,d^{V}\left( {\chi}
\right)}}{{\partial \,\left( {C - C^{B}} \right)}}}\nabla \left(
{C - C^{B}} \right)\,\tilde {C}^{V\times }} \right]} +
{\displaystyle \frac{{S^{E} - G^{E}}}{{C_{eq}^{V\times} } }}  \\
 \\
 -k^{AIkV}\left( {\chi}  \right)\,\,C^{AIk}\,\,\tilde {C}^{V\times}
- k^{IV}\left( {\chi}  \right)\,C_{eq}^{I\times}  \,\tilde
{C}^{I\times }\,\,\tilde {C}^{V\times}  - {\displaystyle
\frac{{S^{V}}}{{C_{eq}^{V\times} } }} +
{\displaystyle \frac{{G^{VT} + G^{VR}}}{{C_{eq}^{V\times} } }} = 0 \\
 \end{array}  \quad {\rm ,}
\end{equation}

\textbf{2) equation for diffusion of silicon self-interstitial}

\begin{equation} \label{Self-interstitial}
\begin{array}{l}
 \nabla {\left[ {d^{I}\left( {\chi}  \right)\,\nabla \tilde {C}^{I\times} }
\right]} - \nabla {\left[ {\left( {\omega ^{\chi}  - 1}
\right){\displaystyle \frac{{\partial \,d^{I}\left( {\chi}
\right)}}{{\partial \,\left( {C - C^{B}} \right)}}}\nabla \left(
{C - C^{B}} \right)\,\tilde {C}^{I\times }} \right]} +
{\displaystyle \frac{{S^{F} - G^{F}}}{{C_{eq}^{I\times} } }} \\
 \\
 -k^{W}\left( {\chi}  \right)\,\,\tilde {C}^{I\times}  +
{\displaystyle \frac{{k^{AIk}}}{{C_{eq}^{I\times} } }}C^{AIk} -
k^{IV}\left( {\chi} \right)\,C_{eq}^{V\times}  \,\tilde
{C}^{V\times} \,\,\tilde {C}^{I\times} - {\displaystyle
\frac{{S^{I}}}{{C_{eq}^{I\times} } }} + {\displaystyle
\frac{{G^{IT} +
G^{IR}}}{{C_{eq}^{I\times} } }} = 0 \\
 \end{array} \quad {\rm ,}
\end{equation}

\noindent where

\begin{equation} \label{Omega}
\omega ^{\chi}  = {\frac{{\chi} }{{k_{B} T}}}{\frac{{\partial
\,\mu ^{\chi }}}{{\partial \,\chi} }} \, {\rm .}
\end{equation}

Here $\tilde{C}^{V\times}$ and $\tilde{C}^{I\times}$ are
respectively the concentrations of nonequilibrium vacancies and
silicon self-interstitials in the neutral charge state, normalized
to the equilibrium concentrations of these species
$\tilde{C}^{V\times}_{eq}$ and $\tilde{C}^{I\times}_{eq}$; $C$ and
$C^{B}$ are the concentrations of substitutionally dissolved
impurity and impurity with the opposite-type of conductivity,
respectively; $C^{AIk}$ is the concentration of impurity
interstitials with the charge state $k$; $\chi$ is the
concentration of charge carriers (electrons $n$ or holes $p$ for
doping with donor or acceptor impurities, respectively) normalized
to the intrinsic concentration of charge carriers $n_{i}$;
$\omega^{\chi}$ is the function which describes a deviation of an
electron (hole) system beyond the perfect solubility; $\mu^{\chi}$
is the chemical potential of electrons (holes); $d^{V}(\chi)$ is
the effective diffusivity of vacancies; $k^{AIkV}$ and $k^{IV}$
are respectively the effective recombination coefficients of
impurity interstitials (in a charge state $k$) and silicon
self-interstitials with vacancies; $G^{E}$ and $S^{E}$ are the
rates of generation and dissolution of the ``impurity atom --
vacancy'' pairs; $S^{V}$ is the rate of the trapping of the
vacancies on the immobile imperfections of a crystalline lattice;
$G^{VT}$ and $G^{VR}$ are the rates of thermal generation of
vacancies and generation of the vacancies due to external
irradiation; $d^{I}(\chi)$ is the effective diffusivity of silicon
self-interstitials; $k^{W}$ is the effective coefficient of the
replacement of the impurity atom by self-interstitial from the
substitutional position into the interstitial one (Watkins effect
\cite{Watkins-69}); $k^{AIk}$ is the effective coefficient for
conversion of impurity atoms from an interstitial to the
substitutional position (phenomenon opposite to the Watkins
effect); $G^{F}$ and $S^{F}$ are the rates of generation and
dissolution of the ``impurity atom -- silicon self-interstitial''
pairs; $S^{I}$ is the rate of the trapping of the silicon
self-interstitials on the immobile sinks of a crystalline lattice;
$G^{IT}$ and $G^{IR}$ are the rates of a thermal generation of
silicon self-interstitials and their generation due to external
irradiation.

The diffusion equations obtained have the following characteristic
features:

\textbf{(i)} these two equations describe diffusion of all point
defects with different charge states as a whole, although only the
concentration of the neutral vacancies $\tilde{C}^{V\times}$ and
silicon self-interstitials $\tilde{C}^{I\times}$ must be derived
to solve equations (\ref{Vacancy}) and (\ref{Self-interstitial}),
respectively. After the solution, the distributions of charged
species, namely, vacancies in a charge state $r$ and silicon
self-interstitials in a charge state $q$, can be calculated from
the expressions describing the local thermodynamic equilibrium
$C^{Vr}=\tilde{C}^{V\times}C^{V\times}_{eq}h^{Vr}\chi^{-\displaystyle
zz^{Vr}}$ and
$C^{Iq}=\tilde{C}^{I\times}C^{I\times}_{eq}h^{Iq}\chi^{-\displaystyle
zz^{Iq}}$. Here $z$, $z^{Vr}$, and $z^{Ir}$ are respectively the
charge of a substitutional impurity atom, the charge of a vacancy
in the charge state $r$, and the charge of a silicon
self-interstitial in the charge state $q$ in terms of the
elementary charge; $h^{Vr}$ and $h^{Iq}$ are the constants of the
mass action law for reactions of defects conversion from neutral
to nonzero charge states;

\textbf{(ii)} the equations obtained take into account the drift
of all charged species due to the built-in electric field. At the
same time, there is no explicit term that would describe the drift
and be proportional to the first derivative of the concentration
of mobile species that essentially complicates the numerical
solution. To exclude this term, a system of equations describing
diffusion of intrinsic point defects in each charge state was
written. Then, the special mathematical transformations of these
equations were performed using the mass action law for conversions
between different charge states of vacancies and
self-interstitials. As a result of these transformations, the
drift of vacancies and silicon self-interstitials in the electric
field are taken into account in the effective diffusion
coefficients $d^{V}(\chi)$ and $d^{I}(\chi)$;

\textbf{(iii)} the effective diffusion coefficients $d^{V}(\chi)$
and $d^{I}(\chi)$ as well as the effective coefficients of
quasichemical reactions $k^{W}(\chi)$, $k^{AIkV}(\chi)$, and
$k^{IV}(\chi)$ are smooth and monotone functions of the
concentration of dopant atoms.

It is to be noted that equations (\ref{Vacancy}) and
(\ref{Self-interstitial}) are very convenient for numerical
solution and studying the fundamentals of diffusion processes
owing to the features (i), (ii), and (iii). In addition, it
follows from these equations that for defect diffusion in
intrinsic or homogeneously doped semiconductor all nonlinear
coefficients are converted into constants. Then, equations
(\ref{Vacancy}) and (\ref{Self-interstitial}) can be presented for
a one-dimensional (1D) domain in the form

\begin{equation} \label{Point defect}
d_{i} \,{\frac{{d^{\, 2}\,\tilde {C}^{\times} }}{{d\,x^{2}}}} -
{\frac{{\tilde {C}^{\times} }}{{\tau} }}\, + {\frac{{G^{T} +
G^{R}}}{{C_{eq}^{\times} } }} = 0 \quad {\rm ,}
\end{equation}

\noindent where $d_{i}$ and $\tau$ are the diffusivity and the
average lifetime of point defects in intrinsic silicon (we do not
concretize the defect species).

In a number of cases concerning the impurity and point defect
diffusion, it is possible to neglect  the mutual interactions of
vacancies and interstitial atoms. For example, under a
low-temperature oxidation of the surface of a semiconductor,
silicon self-interstitials are the dominating defects in a silicon
crystal \cite{Antoniadis-78}. Therefore, one can neglect
calculation of vacancy distribution in modeling the processes of
impurity diffusion due to negligible vacancy concentration. In
this case, the average lifetime of other defects (silicon
self-interstitials) can be assumed to be constant
$\tau=\tau_{i}=const$. Here $\tau_{i}$ is the average lifetime of
defects in an intrinsic semiconductor under equilibrium
conditions. Using the quantity of the average migration length of
point defects $l_{i}=\sqrt{d_{i}\tau_{i}}$, one can present the
equation of diffusion (\ref{Point defect}) in the following form:

\begin{equation} \label{Point defect norm}
{\frac{{d^{\, 2}\,\tilde {C}^{\times} }}{{d\,x^{2}}}} -
{\frac{{1}}{{l_{i}^{2} }}}\,\tilde {C}^{\times}  + {\frac{{1 +
\tilde {g}\left( {x,t} \right)}}{{l_{i}^{2}} }} = 0 \quad {\rm ,}
\end{equation}

\noindent where $\tilde {g}\left( {x,t} \right)=G^{R}/G^{T}$
represents the generation rate of point defects under
consideration in the volume unit of a semiconductor normalized to
the thermal generation rate of these defects.

Let us obtain a solution of equation (\ref{Point defect norm}) in
the 1D finite-length domain $[0,x_{B}]$ for $\tilde {g}\left(
{x,t} \right)=\tilde {g}\left( {x} \right)$ and the Robin boundary
conditions on the left and right boundaries:

\begin{equation} \label{RobinLeft}
- {\left. {w_{1}^{S} d_{i}{\frac{{d\,\tilde {C}^{\times}
}}{{d\,x}}}} \right|_{\displaystyle x = 0}}+ w_{2}^{S} \tilde
{C}^{\times}(0) = w_{3}^{S} \quad {\rm ,}
\end{equation}

\begin{equation} \label{RobinRight}
- {\left. {w_{1}^{B} d_{i}{\frac{{d\,\tilde {C}^{\times}
}}{{d\,x}}}} \right|_{\displaystyle x = x_{B}}}+ w_{2}^{B} \tilde
{C}^{\times}(x_{B}) = w_{3}^{B} \quad {\rm ,}
\end{equation}

\noindent where $w_{1}^{S}$, $w_{2}^{S}$, $w_{3}^{S}$,
$w_{1}^{B}$, $w_{2}^{B}$, and $w_{3}^{B}$ are the constant
coefficients specifying the concrete type of real boundary
conditions.

\section{Solution of the equation describing defect diffusion}
For the solution of the boundary-value problem (\ref{Point defect
norm}), (\ref{RobinLeft}), and (\ref{RobinRight}) we can use the
Green function approach \cite{Butkovskiy-79}:

\begin{equation} \label{Integral}
\tilde {C}^{\times} (x,t) = {\int\limits_{0}^{x_{B}}  {G(x,\xi
)\,\omega \,(\xi )\,d\xi} }  \quad {\rm ,}
\end{equation}

\noindent where the standardizing function $\omega(\xi)$ has the
following form:

\begin{equation} \label{Sfuction}
\omega(\xi)={\frac{{1 + \tilde {g}\left( {x,t}
\right)}}{{l_{i}^{2}} }}+\omega_{S}(\xi)+\omega_{B}(\xi)
\end{equation}

\noindent and $G(x,\xi )$ is the Green's function for equation
(\ref{Point defect norm}). Using the standardizing function
$\omega(\xi)$ \cite{Butkovskiy-79} allows one to reduce the
previous boundary-value problem to the boundary-value problem with
boundary conditions having zero right hand sides:

\begin{equation} \label{RobinLeftZ}
- {\left. {w_{1}^{S} d_{i}{\frac{{d\,\tilde {C}^{\times}
}}{{d\,x}}}} \right|_{\displaystyle x = 0}}+ w_{2}^{S} \tilde
{C}^{\times}(0) = 0 \quad {\rm ,}
\end{equation}

\begin{equation} \label{RobinRightZ}
- {\left. {w_{1}^{B} d_{i}{\frac{{d\,\tilde {C}^{\times}
}}{{d\,x}}}} \right|_{\displaystyle x = x_{B}}}+ w_{2}^{B} \tilde
{C}^{\times}(x_{B}) = 0 \quad {\rm .}
\end{equation}

The Green function for equation (15) with boundary conditions (13)
and (14) has the following form \cite{Butkovskiy-79}:

\begin{equation} \label{Green}
G(x,\xi ) = {\frac{{1}}{{K}}}{\left\{ {\begin{array}{l}
 {Q_{\,1} (x)\;Q_{2} (\xi )\quad \quad 0 \leqslant x \leqslant \xi \leqslant x_{B} } \\
 {Q_{\,1} (\xi )\;Q_{2} (x)\quad \quad 0 \leqslant \xi \leqslant x \leqslant x_{B} } \\
 \end{array}} \right.} \quad {\rm ,}
\end{equation}

\noindent where

\begin{equation} \label{GreenK}
K = - [Q_{\;1} (x){Q}'_{2} (x) - Q_{2} (x){Q}'_{\;1} (x)] = const
\quad {\rm .}
\end{equation}

Here $Q_{1}(x)$ and $Q_{2}(x)$ are the linearly independent
solutions of the homogeneous equation

\begin{equation} \label{Homogeneous}
{\frac{{d^{\, 2}\tilde {C}^{\times} }}{{dx^{2}}}} -
{\frac{{1}}{{l_{i}^{2} }}}\tilde {C}^{\times}  = 0 \quad {\rm .}
\end{equation}

\noindent with the following conditions on the left boundary:

\begin{equation} \label{BoundaryLeft}
Q_{\;1} \left( {0} \right) = - w_{1}^{S} d_{i} \, \, , \quad
{Q}'_{\;1} \left( {0} \right) = - w_{2}^{S}
\end{equation}

\noindent and on the right one:

\begin{equation} \label{BoundaryRight}
Q_{\;2} \left( {x_{B}}  \right) = - w_{1}^{B} d_{i} \, \, , \quad
{Q}'_{\;2} \left( {x_{B}} \right) = - w_{2}^{B} {\rm .}
\end{equation}

Taking into account \cite{Butkovskiy-79}, we can write the
functions $\omega _{S} (x)$ and $\omega _{B} (x)$ as

\begin{equation} \label{OmegaLeft}
\omega _{S} (x) = {\left\{ {\begin{array}{l}
 {{\displaystyle \frac{{1}}{{w_{1}^{S} d_{i}} }}\delta ( - x)w_{3}^{S} ,\quad если\quad
w_{1}^{S} \ne 0} \\
 {{\displaystyle \frac{{1}}{{w_{2}^{S}} }}{\delta} '( - x)w_{3}^{S} ,\quad если\quad
w_{2}^{S} \ne 0} \\
 \end{array}} \right.} \quad {\rm ,}
\end{equation}

\begin{equation} \label{OmegaRight}
\omega _{B} (x) = {\left\{ {\begin{array}{l}
 { - {\displaystyle \frac{{1}}{{w_{1}^{B} d^{AI}}}}\delta (x_{B} - x)w_{3}^{B} ,\quad
если\quad w_{1}^{B} \ne 0} \\
 { - {\displaystyle \frac{{1}}{{w_{2}^{B}} }}{\delta} '(x_{B} - x)w_{3}^{B} ,\quad
если\quad w_{2}^{B} \ne 0} \\
 \end{array}} \right.} \quad {\rm .}
\end{equation}

Let us consider the following Robin boundary condition on the left
boundary of the layer (for example, on the surface $x=0$) and in
the bulk of a semiconductor $x=x_{B}$:

\begin{equation} \label{BounLeft}
w_{1}^{S} = 1 \quad {\rm ,} \qquad w_{2}^{S} \ne 0 \quad {\rm,}
\qquad w_{3}^{S} = 0 \quad {\rm ,}
\end{equation}

\begin{equation} \label{BounRight}
w_{1}^{B} = 1 \quad {\rm ,} \qquad w_{2}^{B} \ne 0 \quad {\rm,}
\qquad w_{3}^{B} = 0 \quad {\rm .}
\end{equation}

These boundary conditions are very interesting for technology
because they allow one to describe the flux of point defects
through the left and the right boundaries as well as the
absorption of defects on the boundary \cite{Ryssel-86}. It follows
from (\ref{BounLeft}) and (\ref{BounRight}) that $\omega _{S}
(x)=0$ and $\omega _{B} (x)=0$, whereas the solutions $Q_{1}$ and
$Q_{2}$ have the following form:

\begin{equation} \label{Q1}
Q_{1} (x) = - {\frac{{1}}{{2}}}{\left[ {\left( {d_{i} + l_{i}
w_{2}^{S}} \right)\;e^{{\frac{{x}}{{l_{i}} }}} + \left( {d_{i} -
l_{i} w_{2}^{S}} \right)\;e^{ - {\frac{{x}}{{l_{i}} }}}} \right]}
\quad {\rm ,}
\end{equation}

\begin{equation} \label{Q2}
Q_{2} (x) = - {\frac{{1}}{{2}}}{\left[ {\left( {d_{i} - l_{i}
w_{2}^{B}} \right)\;e^{{\frac{{x_{B} - x}}{{l_{i}} }}} + \left(
{d_{i} + l_{i} w_{2}^{B}}  \right)\;e^{ - {\frac{{x_{B} -
x}}{{l_{i}} }}}} \right]} \quad {\rm .}
\end{equation}

Then, the constant $K$ is equal to

\begin{equation} \label{K}
K = - {\frac{{1}}{{2l_{i}} }}{\left[ {\left( {d_{i} - l_{i}
w_{2}^{B}} \right)\;\left( {d_{i} + l_{i} w_{2}^{S}}
\right)\;e^{{\frac{{x_{B} }}{{l_{i}} }}} - \left( {d_{i} + l_{i}
w_{2}^{B}}  \right)\;\left( {d_{i} - l_{i} w_{2}^{S}} \right)\;e^{
- {\frac{{x_{B}} }{{l_{i}} }}}} \right]} \quad {\rm .}
\end{equation}

\noindent and the Green function has the following form:

\begin{equation} \label{GreenRobin}
\begin{array}{l}
 G(x,\xi ) = - \displaystyle {{\frac{{l_{i}} }{{2{\left[ {\left( {d_{i} - l_{i} w_{2}^{B}}
\right)\;\left( {d_{i} + l_{i} w_{2}^{S}} \right)\;e^{{
\frac{{x_{B} }}{{l_{i}} }}} - \left( {d_{i} + l_{i} w_{2}^{B}}
\right)\;\left( {d_{i} - l_{i} w_{2}^{S}} \right)\;e^{ -
{\frac{{x_{B}} }{{l_{i}} }}}} \right]}}}}}
\\
 \\
 \times {\left\{ {\begin{array}{l}
 {{\left[ {\left( {d_{i} + l_{i} w_{2}^{S}}  \right)\;e^{{\frac{{x}}{{l_{i}
}}}} + \left( {d_{i} - l_{i} w_{2}^{S}}  \right)\;e^{ -
{\frac{{x}}{{l_{i}
}}}}} \right]}} \\
 {} \\
 {\times {\left[ {\left( {d_{i} - l_{i} w_{2}^{B}}
\right)\;e^{{\frac{{x_{B} - \xi} }{{l_{i}} }}} + \left( {d_{i} +
l_{i} w_{2}^{B}}  \right)\;e^{ - {\frac{{x_{B} - \xi} }{{l_{i}}
}}}} \right]}\quad \quad
0 \le x \le \xi \le x_{B} \;,} \\
 {} \\
 {{\left[ {\left( {d_{i} + l_{i} w_{2}^{S}}  \right)\;e^{{\frac{{\xi
}}{{l_{i}} }}} + \left( {d_{i} - l_{i} w_{2}^{S}}  \right)\;e^{ -
{\frac{{\xi} }{{l_{i}} }}}} \right]}} \\
 {} \\
 {\times {\left[ {\left( {d_{i} - l_{i} w_{2}^{B}}
\right)\;e^{{\frac{{x_{B} - x}}{{l_{i}} }}} + \left( {d_{i} +
l_{i} w_{2}^{B}}  \right)\;e^{ - {\frac{{x_{B} - x}}{{l_{i}} }}}}
\right]}\quad  \quad 0\le \xi \le x \le x_{B} \;.} \\
 \end{array}} \right.} \\
 \end{array}
\end{equation}

Let us assume that a generation of nonequilibrium point defects
occurs due to ion implantation and that the distribution of their
generation rate is approximated by the Gaussian function:

\begin{equation} \label{Gauss}
\tilde {g}\left( {x,t} \right) = g_{m} \exp {\left[ { -
{\frac{{\left( {x - R_{pd}}  \right)^{2}}}{{2\Delta R_{pd}^{2}}
}}} \right]} \quad ,
\end{equation}

\noindent where $g_{m}$ is the maximum rate of generation of
nonequilibrium defects normalized to the rate of the thermal
generation of this species; $R_{pd}$ is the position of the
generation maximum and $\Delta R_{pd}$ is the standard deviation.

Substituting the Green function (\ref{GreenRobin}) and
expression(\ref{Gauss}) into (\ref{Integral}) allows one to obtain
a spatial distribution of point defect concentration:

\begin{equation} \label{GenSolution}
\tilde {C}^{\times} (x) = \tilde {C}_{eq}^{\times}(x) + \tilde
{C}_{R}^{\times}(x) \quad ,
\end{equation}

\noindent where $\tilde {C}_{eq}^{\times}(x)$ is the distribution
of point defect concentration in the case of zero external
radiation and $\tilde {C}_{R}^{\times}(x)$ is the change of defect
concentration due to ion implantation:

\begin{equation} \label{EqSolution}
\begin{array}{l}
 \tilde {C}_{eq}^{\times}  (x) = {\left\{ {\left( {d_{i} - l_{i} w_{2}^{S}}
\right)\,\left( {d_{i} + l_{i} w_{2}^{B}}  \right) - \left( {d_{i}
+ l_{i} w_{2}^{S}}  \right)\,\left( {d_{i} - l_{i} w_{2}^{B}}
\right)\,e^{{\displaystyle \frac{{2x_{B}} }{{l_{i}} }}}} \right.} \\
 \quad \quad \quad + {\left[ {\left( {d_{i} + l_{i} w_{2}^{B}}
\right)\,l_{i} w_{2}^{S} - \left( {d_{i} + l_{i} w_{2}^{S}}
\right)\,l_{i} w_{2}^{B} e^{{\displaystyle \frac{{x_{B}} }{{l_{i}}
}}}} \right]}\,e^{{\displaystyle \frac{{x}}{{l_{i} }}}}
\\
\\
 \quad \quad \quad {\left. { + {\left[ {\left( {d_{i} - l_{i} w_{2}^{B}}
\right)\,l_{i} w_{2}^{S} \,e^{{\displaystyle \frac{{2x_{B}}
}{{l_{i}} }}} - \left( {d_{i} - l_{i} w_{2}^{S}}  \right)\,l_{i}
w_{2}^{B} e^{{\displaystyle \frac{{x_{B}} }{{l_{i}
}}}}} \right]}\,e^{ - {\displaystyle \frac{{x}}{{l_{i}} }}}} \right\}}
\\
\\
 \quad \quad \quad \times {\left[ {\left( {d_{i} - l_{i} w_{2}^{S}}
\right)\;\left( {d_{i} + l_{i} w_{2}^{B}}  \right) - \left( {d_{i}
+ l_{i}
w_{2}^{S}}  \right)\;} \right]}^{ - 1}\; \\
\end{array}
\end{equation}

\noindent and

\begin{equation} \label{RadSolution}
\tilde {C}_{R}^{\times}  (x) = \tilde {C}_{R1}^{\times}  (x) +
\tilde {C}_{R2}^{\times}  (x) \quad ,
\end{equation}

\noindent where

\begin{equation} \label{Rad1}
\begin{array}{l}
 \tilde {C}_{R1}^{\times}  (x) = g_{m} \sqrt {{\displaystyle \frac{{\pi} }{{2}}}}
{\displaystyle \frac{{\Delta R_{pd}} }{{2l_{i}}
}}e^{{\displaystyle \frac{{\Delta R_{pd}^{2} - 2l_{i} R_{pd} +
2l_{i} x_{B}} }{{2l_{i}^{2}} }}}
\\
\\
\times {\left[ {\left( {d_{i} - l_{i} w_{2}^{B}}
\right)\;e^{{\displaystyle \frac{{x_{B} - x}}{{l_{i}} }}} + \left(
{d_{i} + l_{i} w_{2}^{B}}  \right)\;e^{ - {\displaystyle
\frac{{x_{B} - x}}{{l_{i}} }}}} \right]}
\\
\\
\times {\left\{ {\left( {d_{i} + l_{i} w_{2}^{S}}
\right)\;e^{{\displaystyle \frac{{2R_{pd} }}{{l_{i}} }}}{\left[
{erf\left( {{\displaystyle \frac{{\Delta R_{pd}^{2} + l_{i} R_{pd}
- l_{i} x}}{{\sqrt {2} \;\Delta R_{pd} l_{i}} }}} \right) -
erf\left( {{\displaystyle \frac{{\Delta R_{pd}^{2} + l_{i} R_{pd}}
}{{\sqrt {2} \;\Delta R_{pd}
l_{i}} }}} \right)} \right]}} \right.}
\\
\\
{\left. { + \left( {d_{i} - l_{i} w_{2}^{S}}  \right)\;{\left[
{erf\left( {{\displaystyle \frac{{\Delta R_{pd}^{2} - l_{i}
R_{pd}} }{{\sqrt {2} \;\Delta R_{pd} l_{i}} }}} \right) -
erf\left( {{\displaystyle \frac{{\Delta R_{pd}^{2} - l_{i} R_{pd}
+ l_{i} x}}{{\sqrt {2} \;\Delta R_{pd} l_{i}} }}} \right)}
\right]}} \right\} }
\\
\\
\times {\left[ {\left( {d_{i} - l_{i} w_{2}^{S}}  \right)\;\left(
{d_{i} + l_{i} w_{2}^{B}}  \right) - \left( {d_{i} + l_{i}
w_{2}^{S}} \right)\;\left( {d_{i} - l_{i} w_{2}^{B}}
\right)\;e^{{\displaystyle \frac{{2x_{B}
}}{{l_{i}} }}}} \right]}^{ - 1}\quad , \\
 \end{array}
\end{equation}

\begin{equation} \label{Rad2}
\begin{array}{l}
 \tilde {C}_{R2}^{\times}  (x) = g_{m} \sqrt {{\displaystyle \frac{{\pi} }{{2}}}}
{\displaystyle \frac{{\Delta R_{pd}} }{{2l_{i}}
}}e^{{\displaystyle \frac{{\Delta R_{pd}^{2} - 2l_{i} R_{pd} -
2l_{i} x}}{{2l_{i}^{2}} }}}{\left[ {d_{i} - l_{i} w_{2}^{S} +
\left( {d_{i} + l_{i} w_{2}^{S}} \right)\;e^{{\displaystyle
\frac{{2x}}{{l_{i}} }}}}
\right]} \\
\times  {\left\{ {\left( {d_{i} + l_{i} w_{2}^{B}}
\right)\;e^{{\displaystyle \frac{{2R_{pd} }}{{l_{i}} }}}{\left[
{erf\left( {{\displaystyle \frac{{\Delta R_{pd}^{2} + l_{i} R_{pd}
- l_{i} x_{B}} }{{\sqrt {2} \;\Delta R_{pd} \,l_{i}} }}} \right) -
erf\left( {{\displaystyle \frac{{\Delta R_{pd}^{2} + l_{i} R_{pd}
- l_{i} x}}{{\sqrt {2} \;\Delta
R_{pd} \,l_{i}} }}} \right)} \right]}} \right.} \\
 {\left. { + \left( {d_{i} - l_{i} w_{2}^{B}}  \right)\;e^{{\displaystyle \frac{{2x_{B}
}}{{l_{i}} }}}{\left[ {erf\left( {{\displaystyle \frac{{\Delta
R_{pd}^{2} - l_{i} R_{pd} + l_{i} x}}{{\sqrt {2} \;\Delta R_{pd}
\,l_{i}} }}} \right) - erf\left( {{\displaystyle \frac{{\Delta
R_{pd}^{2} - l_{i} R_{pd} + l_{i} x_{B}} }{{\sqrt {2}
\;\Delta R_{pd} \,l_{i}} }}} \right)} \right]}} \right\}}  \\
 \times {\left[ {\left( {d_{i} - l_{i} w_{2}^{S}}  \right)\;\left( {d_{i} +
l_{i} w_{2}^{B}}  \right) - \left( {d_{i} + l_{i} w_{2}^{S}}
\right)\;\left( {d_{i} - l_{i} w_{2}^{B}}
\right)\;e^{{\displaystyle \frac{{2x_{B}
}}{{l_{i}} }}}} \right]}^{ - 1}\quad . \\
 \end{array}
\end{equation}

It was mentioned above that in the up-to-date electronics
different layered structures such as Ge$_{{\rm x}}$Si$_{{\rm
1}{\rm -} {\rm x}}$/Si or silicon-on-insulator (SOI) are often
used for decreasing the device dimensions and improving the device
performance. The derived analytical solution for a finite-length
domain $[0,x_{B}]$ is convenient for modeling and investigating
point defect diffusion in a separate layer of these structures.
For example, in Fig. 1 the calculated distribution of point
defects in the silicon layer of thickness 0.4 $\mu $m is
presented. Primarily, the case of zero external radiation ($g_{m}$
= 0) is considered for the better understanding of the influence
of ion implantation.

\begin{figure}[!ht]
\centering {
\begin{minipage}[!ht]{9.4 cm}
{\includegraphics[scale=0.8]{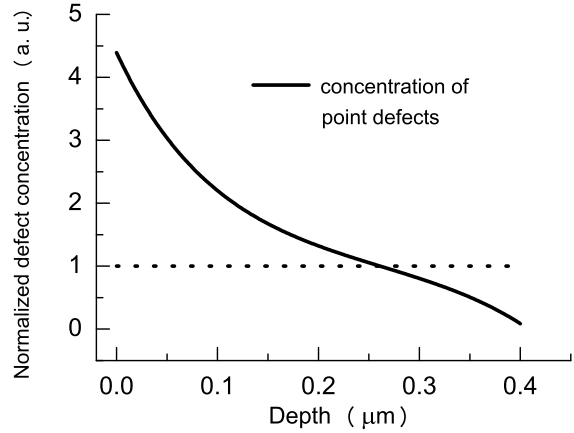}}
\end{minipage}
}

\caption{Calculated concentration distribution of the neutral
point defects in a silicon layer of thickness 0.4 $\mu $m. The
dotted curve represents the thermally equilibrium value of the
normalized concentration of neutral point defects}
\label{fig:Calcequilibrium}
\end{figure}

It is evident that for zero fluxes of defects through the
boundaries of the layer a distribution of point defects is
homogeneous and the value of normalized concentration of these
defects is equal to 1 (dotted curve). Deviation from the uniform
defect distribution occurs only if there are nonzero fluxes of
defects through the boundaries or there is an absorption
(generation) of point defects on the surface or at the interface.
For example, the distribution of defects presented in Fig.
~\ref{fig:Calcequilibrium} is calculated under the assumption that
two fluxes of point defects through the left and right boundaries
are directed along the $x$ axis. With this purpose the
coefficients $w_{2}^{S}$ and $w_{2}^{B}$ have been presented in
the following form:

\begin{equation} \label{EscapeVelocity}
w_{2}^{S} = {\rm v}_{eff}^{S} \quad {\rm ,} \qquad w_{2}^{B} = -
{\rm v}_{eff}^{B} \quad ,
\end{equation}

\noindent where ${\rm v}_{eff}^{S}$ and ${\rm v}_{eff}^{B}$ are
the effective rate of point defect removal outside the layer
through the left and right boundaries, respectively. For defect
distribution presented in Fig. ~\ref{fig:Calcequilibrium}, the
values ${\rm v}_{eff}^{S}$ =-0.0094 $\mu $m/s and ${\rm
v}_{eff}^{B}$ = 4.0 $\mu $m/s were used. Also, the value of the
average migration length of point defects $l_{i}$ = 0.1 $\mu $m
and the value of intrinsic diffusivity $d_{i}$ = 0.01 $\mu
$m$^{{\rm 2}}$/s were chosen. It can be seen from Fig.
~\ref{fig:Calcequilibrium} that according to the boundary
conditions (\ref{EscapeVelocity}) used for solving equation
(\ref{Point defect norm}) the concentration of the point defects
in the vicinity of the left boundary increases due to supplying
additional defects in the layer, whereas near to the right
boundary the concentration of intrinsic point defects decreases
due to the removal of this species outside the layer. The
analytical solution obtained describes the distribution of the
concentration of point defects in the neutral charge state. The
concentration of the charged defect species $C^{r}(x)$ can be
calculated from the above-mentioned expressions
$C^{Vr}=\tilde{C}^{V\times}C^{V\times}_{eq}h^{Vr}\chi^{-\displaystyle
zz^{Vr}}$ and
$C^{Iq}=\tilde{C}^{I\times}C^{I\times}_{eq}h^{Iq}\chi^{-\displaystyle
zz^{Iq}}$ that follow from the mass action law.

It is worth noting that due to the quasi-stationarity of the
diffusion equation for point defects, exactly the same solution
takes place for the Dirichlet boundary conditions with
$\tilde{C}^{\times}(0)=\tilde{C}^{\times}_{S}$= 4.393 a.u. and
$\tilde{C}^{\times}(x_{B})=\tilde{C}^{\times}_{B}$= 0.08689 a.u.
Here $\tilde{C}^{\times}_{S}$ and $\tilde{C}^{\times}_{B}$ are the
normalized concentrations of intrinsic point defects on the left
(surface) and right boundaries of the layer.

Let us consider now the main features of the solutions of equation
(\ref{Point defect norm}) in the case of intense generation of
nonequilibrium point defects in the vicinity of the surface. Such
generation can occur during low-energy implantation of hydrogen
ions into semiconductor substrate. For example, let us suppose
that the energy of hydrogen implantation is 2 keV. Then,
calculation performed by the code SRIM \cite{SRIM-13} gives the
following values: $R_{pd}$ = 0.033 $\mu $m, $\Delta R_{pd}$ =
0.0248 $\mu $m, if one assumes that the distribution of generated
defects is proportional to the distribution of implanted hydrogen
ions.

\begin{figure}[!ht]
\centering {
\begin{minipage}[!ht]{9.4 cm}
{\includegraphics[scale=0.8]{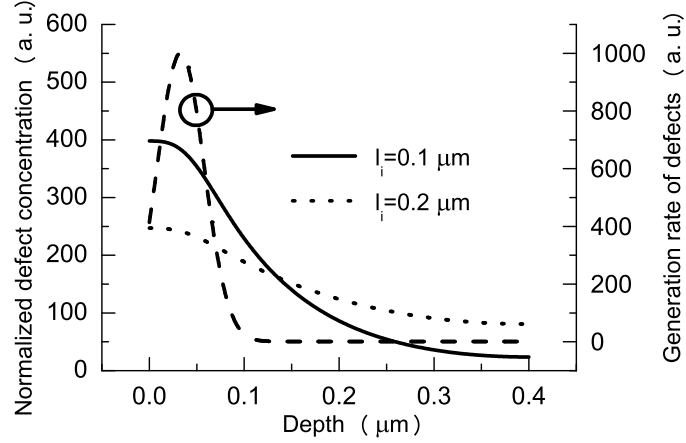}}
\end{minipage}
}

\caption{Concentration distribution of the neutral point defects
normalized to the thermally equilibrium value of defect
concentration in a silicon layer of thickness 0.4 $\mu $m for the
case of hydrogen implantation with an energy of 2 keV. The dashed
line represents the generation rate of point defects normalized to
the equilibrium one} \label{fig:CalImplant}
\end{figure}

In Fig. ~\ref{fig:CalImplant} the calculated concentration
distribution of nonequilibrium point defects in the silicon layer
of thickness 0.4 $\mu $m is presented. It was supposed that the
maximal generation rate of point defects due to the ion
implantation exceeds 1000 times the rate of thermal generation
($g_{m}$ =1000), whereas the diffusion parameters are the same
($l_{i}$ = 0.1 $\mu $m, $d_{i}$ = 0.01 $\mu $m$^{{\rm 2}}$/s). For
comparison, the point defect distribution calculated for the value
$l_{i}$ = 0.2 $\mu $m is also presented. The case of zero fluxes
through the left and right boundaries is investigated primarily.

It follows from Fig. ~\ref{fig:CalImplant} that the point defect
concentration decreases 1.6 times at the surface of a
semiconductor if the average migration length increases 2 times.
Simultaneously, the distribution of point defects becomes flatter.
On the other hand, there is a significant increase of the point
defect concentration, more accurately by a factor of 3.4, on the
right boundary of the layer.

\begin{figure}[!ht]
\centering {
\begin{minipage}[!ht]{9.4 cm}
{\includegraphics[scale=0.8]{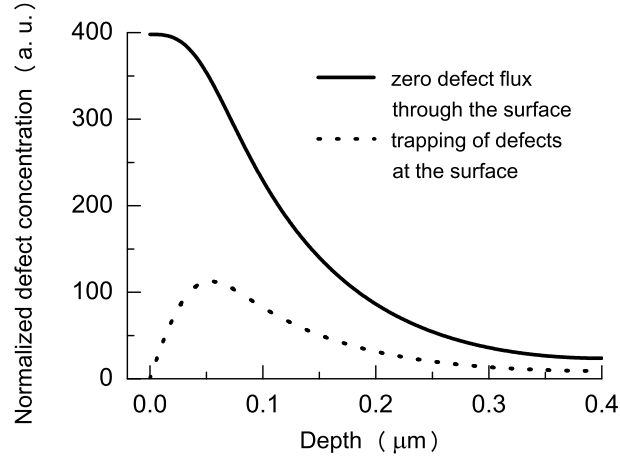}}
\end{minipage}
}

\caption{Concentration distribution of the neutral point defects
normalized to the thermally equilibrium value of defect
concentration in a silicon layer of thickness 0.4 $\mu $m for the
case of hydrogen implantation with an energy of 2 keV. The solid
line represents distribution of point defects calculated for the
case of zero defect flux through the left boundary, whereas the
dotted line describes diffusion of point defects under conditions
of defect trapping on the surface} \label{fig:CalImplant2keV}
\end{figure}

Now, this paper investigates the main features of the solution
obtained for the case of defect removal through the left boundary
of the layer. It was mentioned above that this boundary condition
is also similar to defect trapping on the surface of a
semiconductor. With this purpose Fig. ~\ref{fig:CalImplant2keV}
presents two distributions of defects which were calculated for
the case of zero defect flux through the left boundary and for the
case of intensive trapping of defects by the surface,
respectively. It is supposed that the average migration length of
point defects is equal to 0.1 $\mu $ m. It can be seen from Fig. 3
that the trapping of point defects on the surface results in the
change of the form of its concentration profile and in the
significant decrease of defect concentration. For example, the
maximal concentration of point defects decreases 3.5 times.

\begin{figure}[!ht]
\centering {
\begin{minipage}[!ht]{9.4 cm}
{\includegraphics[scale=0.8]{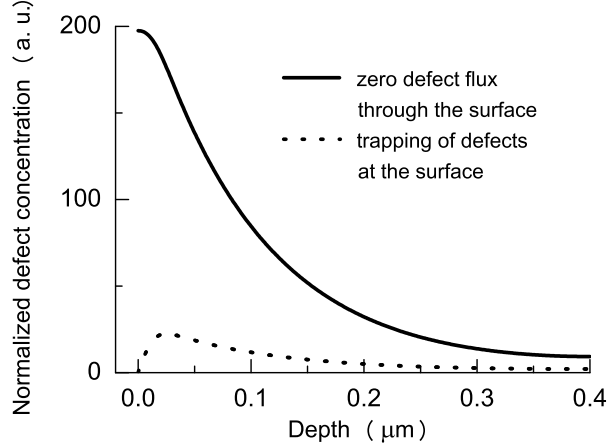}}
\end{minipage}
}

\caption{Concentration distribution of the neutral point defects
normalized to the thermally equilibrium value of defect
concentration in a silicon layer of thickness 0.4 $\mu $m for the
case of hydrogen implantation with an energy of 500 eV. The solid
line represents distribution of point  defects calculated for the
case of zero defect flux through the left boundary, whereas the
dotted line describes diffusion of point defects under conditions
of defect trapping on the surface} \label{fig:CalImplant500eV}
\end{figure}

More serious influence of the surface on the distribution of point
defects can be observed for small values of implantation energy.
It can be seen from Fig. ~\ref{fig:CalImplant500eV}, where a
similar calculation for the energy of implantation of hydrogen
ions equals to 500 eV is presented. For this value of hydrogen
implantation energy the calculation of the parameters describing
the distribution of implanted ions gives the following values:
$R_{pd}$ = 0.0097 $\mu $m, $\Delta R_{pd}$ = 0.011 $\mu $m
\cite{SRIM-13}.

It can be seen from Fig. ~\ref{fig:CalImplant500eV} that the
maximal concentration of point defects decreases 8.8 times due to
the trapping of point defects on the surface, which is in close
vicinity (a few nanometers) to the region of intense generation of
nonequilibrium point defects.

\newpage

\section{Conclusions}

The analytical solution of the one-dimensional equation that
describes quasi-stationary diffusion of intrinsic point defects in
semiconductor crystals has been obtained for the case of the Robin
boundary conditions on the left and right boundaries of the layer.
It is supposed that the generation rate of nonequilibrium point
defects is approximated by the Gaussian function. To derive an
analytical solution of this boundary-value problem, the Green
function approach has been used.

The solution obtained is focused on the application in modeling
technological processes used for fabrication of modern silicon
integrated microcircuits and other semiconductor devices which
have layered structure. For example, it can be helpful for
verification of the numerical solutions obtained and for
investigation of the features of transport processes of vacancies
and silicon self-interstitial atoms depending upon the
implantation parameters and parameters of boundary conditions. It
follows from a large uncertainty of diffusivity and other
transport properties of point defects known from the literature
that the analytical solution obtained can successfully replace the
numerical solution in modeling a number of technological processes
used in the modern microelectronics.

To illustrate the  usefulness of the obtained solution, the
investigation of the changes in the form and concentration values
of distribution of point defects has been carried out for
different boundary conditions and two values of the average
migration length of diffusing species. The cases of pure thermal
generation of point defects within the limits of the layer and
generation of nonequilibrium defects due to hydrogen ion
implantation have been investigated. It has been shown that there
is a strong influence of the surface on the concentration values
and the form of distribution of nonequilibrium point defects when
the implantation energy decreases.


\newpage

\begin{thebibliography}{xxxx}

\bibitem{Pankove-91}
Hydrogen in Semiconductors. Semiconductors and Semimetals. Vol.
34. Volume Editors: J. I. Pankove, N. M. Johnson (Academic Press,
Inc., Harcourt Brace Jovanovich, Publishers, 1991) 629 pages.

\bibitem{Zhang-02}
Yi Zhang, Modeling hydrogen diffusion for solar cell passivation
and process optimization {\it Ph.D. theses, New Jersey Institute
of Technology and Rutgers, the State University of New
Jersey-Newark} (2002).

\bibitem{Chao-05}
D. S. Chao, D. Y. Shu, S. B. Hung, W. Y. Hsieh, and M.-J. Tsai,
Investigation of silicon-on-insulator (SOI) substrate preparation
using the smart-cut$^{TM}$ process, Nuclear Instrum. and Methods
in Phys. Res. B. Vol.237. pp.197Ц202 (2005).

\bibitem{Panteleev-76}
V. A. Panteleev, S. N. Ershov, V.V. Chernyakhovskii, and S. N.
Nabornykh, Determination of the migration of vacancies and of
intrinsic interstitial atoms in silicon in the temperature
interval 400-600 K, JETP Lett. Vol.23, No.12. pp.633-635 (1976).

\bibitem{Hallen-98}
A. Hall\'{e}n, N. Keskitalo, and B. G. Svensson, Diffusion and
reaction kinetics of fast-ion-induced point defects studied by
deep level transient spectroscopy, Defect and Diffusion Forum.
Vols.153-155. pp.193-204 (1998);
DOI:10.4028/www.scientific.net/DDF.153-155.193

\bibitem{Minear-72}
R. L. Minear, D. C. Nelson, and J. F. Gibbons, Enhanced diffusion
in Si and Ge by light ion implantation, J. Appl. Phys. Vol.43,
No.8. pp.3468-3480 (1972).

\bibitem{Ryssel-86}
H. Ryssel, I. Ruge. {\it Ion Implantation} (Wiley, Chichester,
1986) 478 pages.

\bibitem{Portavoce-04}
A. Portavoce, I. Berbezier, P. Gas, and A. Ronda, Sb surface
segregation during epitaxial growth of SiGe heterostructures: The
effects of Ge composition and biaxial stress, Phys. Rev. B.
Vol.69. Art.No.155414 (2004).

\bibitem{Leitz-06}
C. W. Leitz, C. J. Vineis, J. Carlin, J. Fiorenza, G. Braithwaite,
R. Westhoff, V. Yang, M. Carroll, T. A. Langdo, K. Matthews, P.
Kohli, M. Rodder, R. Wise, and A. Lochtefeld, Direct regrowth of
thin strained silicon films on planarized relaxed
silicon--germanium virtual substrates, Thin Solid Films. Vol.513.
pp.300-306 (2006).

\bibitem{Bhandari-08}
J. Bhandari, M. Vinet, T. Poiroux B. Previtali, B. Vincent,L.
Hutin, J. P. Barnes, S. Deleonibus, and A. M. Ionescu, Study of n-
and p-type dopants activation and dopants behavior with respect to
annealing conditions in silicon germanium-on-insulator (SGOI),
Mat. Sci. Eng. B. Vols.154-155. pp.114-117 (2008).

\bibitem{Pichler-04}
P. Pichler, In: {\itshape Computational Microelectronics,
Intrinsic point defects, impurities, and their diffusion in
solids,} edited by S. Selberherr (Springer, Wien, New-York, 2004)
467 pages.

\bibitem{Velichko-84}
O. I. Velichko, ``A set of equations of radiation-enhanced
diffusion of ion-implanted impurities'', in: I. I. Danilovich, A.
G. Koval', V. A. Labunov et al. (Eds.), Proceedings of VII
International Conference ``Vzaimodeistvie Atomnyh Chastits s
Tverdym Telom (Interaction of Atomic Particles with Solid)'', Part
2, Minsk, Belarus, 180-181 (1984) (in Russian).

\bibitem{Velichko-88}
O. I. Velichko. {\it Atomic Diffusion Processes under
Nonequilibrium State of the Components in a Defect--Impurity
System of Silicon Crystals}. Ph.D. Dissertation. (Institute of
Electronics of the National Academy of Sciences of Belarus, Minsk,
1988) (In Russian).

\bibitem{Watkins-69}
G. D. Watkins, A microscopic view of radiation damage in
semiconductors using EPR as a probe. IEEE  Trans. Vol.NS-16, No.6.
pp.13-18 (1969).

\bibitem{Antoniadis-78}
D. A. Antoniadis, A. M. Lin, and R. W. Dutton, Oxidation-enhanced
diffusion of arsenic and phosphorus in near-intrinsic (100)
silicon. Appl. Phys. Lett. Vol.33, No.12. pp.1030-1033 (1978).

\bibitem{Butkovskiy-79}
A. G. Butkovskiy, {\it Harakteristiki system s raspredelennymi
parametrami (Characteristics of Distributed-Parameter Systems)}
(Nauka, Moscow, 1979) (in Russian).

\bibitem{SRIM-13}
J. F. Ziegler, J. P. Biersack, and M. D. Ziegler, SRIM (Stopping
and Range of Ions in Matter) www.SRIM.org

\end{thebibliography}
\end{document}